\documentclass[aps,twocolumn,pre,footinbib,floatfix,showpacs]{revtex4}
\usepackage{epsf,color,colordvi}
\usepackage{graphicx}

\begin{document}
\title{Accurate numerical solutions of the time-dependent 
Schr\"odinger equation}

\author{W. van Dijk}
\email{vandijk@physics.mcmaster.ca}
\affiliation{Physics Department, Redeemer University College, 
  Ancaster, Ontario L9K 1J4, Canada}
\affiliation{Department of Physics and Astronomy, McMaster University, 
  Hamilton, Ontario L8S 4M1, Canada} 
\author{F. M. Toyama}
\email{toyama@cc.kyoto-su.ac.jp}
\affiliation{Department of Information and Communication Sciences, 
  Kyoto Sangyo University, Kyoto 603-8055, Japan}

\begin{abstract}
  We present a generalization of the often-used Crank-Nicolson (CN)
  method of obtaining numerical solutions of the time-dependent
  Schr\"odinger equation.  The generalization yields numerical
  solutions accurate to order $(\Delta x)^{2r-1}$ in space and
  $(\Delta t)^{2M}$ in time for any positive integers $r$ and $M$,
  while CN employ $r=M=1$.  We note dramatic improvement in the
  attainable precision (circa 10 or greater orders of magnitude) along
  with several orders of magnitude reduction of computational time.
  The improved method is shown to lead to feasible studies of
  coherent-state oscillations with additional short-range
  interactions, wavepacket scattering, and long-time studies of
  decaying systems.

\end{abstract}

\date{\today} 


\maketitle

\section{Introduction}
\label{intro}
Whereas there are a number of examples of exact analytic solutions of
time-independent problems in quantum mechanics, such solutions of
time-dependent problems are few.  The analytic solutions of both types
that do exist tend to provide approximate models to actual physical
systems.  No doubt such solutions are instructive for gaining insight
into the behavior of the physical systems that they describe.
Nevertheless because the models themselves are often approximations
and because one wishes to describe real systems as precisely as
possible, one also relies on accurate numerical methods.

The time-dependence of nonrelativistic quantum systems, which is the
focus of this paper, has become important in diverse areas of atomic
and subatomic physics. Examples of these include the study of nuclear
processes such as the decay of unstable nuclei and associated
phenomena like atomic ionization~\cite{vandijk99a, kataoka00} and
bremsstrahlung~\cite{bertulani99, vandijk03a, misicu01}, the study of
fundamental processes necessary for quantum computing~\cite{cheon04},
the study of mesoscopic physics or nanophysics
devices~\cite{veenstra04}, and the motion of atoms in a trap.  A
reliable and accurate numerical determination of the time-dependent
wave function such as we discuss in this paper will no doubt be
necessary and/or helpful in making advances in the understanding of a
variety of quantum processes.

In this paper we consider the numerical solution of the time-dependent
nonrelativistic Schr\"odinger equation.  Much has been learned about
basic scattering processes from the numerically generated solutions of
traveling wavepackets as they pass through a potential
region~\cite{goldberg67}, as well as the time-evolution of unstable
quantum processes~\cite{winter61}.  However, the methods used in the
past, and still employed currently, are limited in that the solutions
often degrade after a certain time interval, so that they reduce to
noise.  Furthermore for processes in which the wave function spreads
or travels away from the source one often requires such a large number
of space steps that the computation becomes prohibitive.

The goal of this study is to improve the existing standard approach by
allowing for relatively large step sizes both in time and in space and
thus to reduce the number of basic arithmetical calculations while
obtaining more accurate solutions.  We have been able to make
significant improvement to one conventional approach, \emph{viz.}, the
Crank-Nicolson (CN) implicit integration scheme for the time-dependent
Schr\"odinger equation.  Many years ago the CN approach was shown to
be successful in the study of wavepacket scattering in one dimension
by Goldberg \emph{et al.}~\cite{goldberg67}.  In recent years the CN
method continues to be employed for its space- and/or time-development
algorithm to study various time-dependent problems.  See, for example,
Refs.~\cite{press_C92,patriarca94,qian06,vyas06}.  The attractive
aspect of this method is that the solution is constrained to be
unitary at every time step.  It is this constraint that makes the
solution stable regardless of time- or space-step size.  Although the
evolution of the solution is unitary, the wave function is not correct
if the step sizes are too large.  The error is of ${\cal O}({(\Delta
  x)^2,(\Delta t)^3})$, where $\Delta x$ and $\Delta t$ are the
spatial and temporal step size, respectively.

The method was successfully generalized to two dimensional scattering
by Galbraith \emph{et al.}~\cite{galbraith84} and, more recently, to
multichannel scattering~\cite{vandijk03b}.  Furthermore alternative
methods which are fast computationally were introduced by Kosloff and
Kosloff~\cite{kosloff83}.  These involve the fast Fourier transform of
the kinetic energy operator of the Schr\"odinger equation.  Variants
of this method were discussed in Ref.~\cite{leforestier91}.  Although
this approach is fast and is able to handle large time intervals in
one pass, it is not unitary and does require a large number of space
intervals.

Improved CN algorithms have been discussed by a few authors.
Mi\c{s}ucu \emph{et al.}~\cite{misicu01} introduced a seven-point
formula for the second-order spatial derivative with error of ${\cal
  O}((\Delta x)^6)$ and an improved time-integration scheme with an
error of ${\cal O}((\Delta t)^5)$.  They claim to obtain two orders of
magnitude improvement in the time advance.  Moyer~\cite{moyer04} uses
a Numerov scheme for the spatial integration method with error of
${\cal O}((\Delta x)^6)$ but has a one-stage time evolution giving the
CN precision in time, i.e., with an error of ${\cal O}((\Delta t)^3)$.
Moyer also introduces transparent boundary conditions for unconfined
systems.  We have found these very useful for making long-time or
large-space problems tractable~\cite{veenstra04}, but we will not
discuss such boundary conditions further in this paper.  Puzynin
\emph{et al.}~\cite{puzynin99,puzynin00} indicate how to generalize
the time development to higher order, but do not discuss spatial
integration.

In this paper we present a generalization of the often-used CN method
of obtaining numerical solutions of the time-dependent Schr\"odinger
equation.  The generalization yields numerical solutions accurate to
order $(\Delta x)^{2r-1}$ in space and $(\Delta t)^{2M}$ in time for
any positive integers $r$ and $M$, while CN employ $r=M=1$.  By
appropriate choice of $r$ and $M$ the improvement can be of such a
nature that hitherto computationally unfeasible problems become
doable, and solutions with low to modest precision can now be obtained
extremely accurately.

In the following we consider the generalization of spatial integration
in Sec.~\ref{spatial} and the generalization of the time integration
in Sec.~\ref{time}.  In Sec.~\ref{errors} we discuss errors and a way
of dealing with a particular type of boundary condition.  We study
specific examples to illustrate the improvement of the generalizations
over the standard CN procedure in Sec.~\ref{examples}.  Some general
observations and conclusions are made in Sec.~\ref{conclusions}.

\section{Spatial integration}
\label{spatial}
\setcounter{equation}{0} We describe a general procedure for solving
the one-dimensional time-dependent Schr\"odinger equation
\begin{equation}\label{eq:2.1}
\left(i\hbar\frac{\partial ~}{\partial t} -
H\right)\psi(x,t)=0,\hspace{.2in} \psi(x,t_0) = \phi(x),
\label{1}
\end{equation}
where 
\begin{equation}\label{eq:2.2}
H = -\frac{\hbar^2}{2m}\frac{\partial^2~}{\partial x^2} + V(x),
\end{equation}
and $\phi(x)$ is a given wave function at initial time $t_0$.  In this
section we use the standard time-advance procedure of the CN
method, but generalize the spatial integration.  In Sec.~\ref{time} we
generalize the time-evolution procedure.

The time evolution of the system can be expressed in terms of an
operator acting on the wave function at time $t$ which gives the wave
function at a later time $t+\Delta t$ according to the equation
\begin{equation}\label{eq:2.3}
  \psi(x,t+\Delta t) = e^{\textstyle -iH\Delta t/\hbar}\psi(x,t).
\end{equation}
\mbox{} \\
The time-evolution operator $e^{\textstyle -iH\Delta t/\hbar}$ can be
expanded to give a unitary approximation of the operator by setting
\begin{equation}\label{eq:2.4}
e^{\textstyle -iH\Delta t/\hbar} = \frac{1 - {1\over 2}{iH\Delta t/\hbar}}
{1 + {1\over 2}{iH\Delta t/\hbar}} + {\cal O}((\Delta t)^3).
\end{equation}
Inserting the approximate form of the operator into
Eq.~(\ref{eq:2.3}), we obtain the equation
\begin{equation}\label{eq:2.5}
  \left(1+{1\over 2}iH\Delta t/\hbar\right)\psi(x,t+\Delta t) =  
  \left(1-{1\over 2}iH\Delta t/\hbar\right)\psi(x,t),
\end{equation}
with an error of ${\cal O}((\Delta t)^3)$.  Here we focus on the
second-order spatial derivative in $H$ of Eq.~(\ref{eq:2.2}) and leave
improvements with respect to the time derivative to Sec.~\ref{time}.
We generalize the usual three-point formula and the seven-point
formula of Mi\c{s}icu \emph{et al.}~\cite{misicu01}, for the
second-order derivative to a $(2r+1)$-point formula.  Such a formula
has the form
\begin{equation}\label{eq:2.6}
 y''(x)\equiv y^{(2)} = \frac{1}{h^2} \sum_{k=-r}^{k=r} c_k^{(r)} \,
  y(x+kh) + {\cal O}(h^{2r}),
\end{equation}
where $c_k^{(r)}$ are real constants.  To obtain the coefficients
$c_k^{(r)}$ we make expansions
\begin{widetext}
\begin{eqnarray}
y(x+kh) & = & y(x) + (kh)y^{(1)}(x) + \frac{1}{2!}(kh)^2 y^{(2)}(x) + \cdots + 
\frac{1}{(2r+1)!}(kh)^{2r+1} y^{(2r+1)}(x) + {\cal O}(h^{2r+2}) \nonumber \\
y(x-kh) & = & y(x) - (kh)y^{(1)}(x) + \frac{1}{2!}(kh)^2 y^{(2)}(x) - \cdots + 
\frac{(-1)^{2r+1}}{(2r+1)!} (kh)^{2r+1}y^{(2r+1)}(x) + {\cal
O}(h^{2r+2}), \nonumber
\end{eqnarray}
for $k = 1,2,\dots,r$;  $y^{(i)}$ denotes the $i$th derivative with
respect to $x$.  When we add the two equations, the terms with
odd-order derivatives cancel, resulting in the equation
\begin{equation}\label{eq:2.7}
2\frac{(kh)^2}{2!}y^{(2)}(x) + 2\frac{(kh)^4}{4!}y^{(4)}(x) + \cdots +
2\frac{(kh)^{2r}}{(2r)!}y^{(2r)}(x) = y(x+kh) + y(x-kh) - 2y(x) +{\cal
O}(h^{2r+2}).
\end{equation}
Thus we obtain the system of $r$ equations in $r$ unknowns, i.e.,
$y^{(2k)}(x)$ for $k=1,\dots,r$,
\begin{eqnarray}
2\frac{(h)^2}{2!}y^{(2)}(x) + 2\frac{(h)^4}{4!}y^{(4)}(x) + \cdots +
2\frac{(h)^{2r}}{(2r)!}y^{(2r)}(x) & = & y(x+h) + y(x-h) - 2y(x) 
\nonumber \\
2\frac{(2h)^2}{2!}y^{(2)}(x) + 2\frac{(2h)^4}{4!}y^{(4)}(x) + \cdots +
2\frac{(2h)^{2r}}{(2r)!}y^{(2r)}(x) & = & y(x+2h) + y(x-2h) - 2y(x) 
\nonumber \\ 
\vdots\hspace{2.in} && \nonumber \\
2\frac{(rh)^2}{2!}y^{(2)}(x) + 2\frac{(rh)^4}{4!}y^{(4)}(x) + \cdots +
2\frac{(rh)^{2r}}{(2r)!}y^{(2r)}(x) & = & y(x+rh) + y(x-rh) - 2y(x). 
\label{eq:2.8}
\end{eqnarray}
\end{widetext}

We solve these equations to obtain $y^{(2)}(x)$.  It is evident from
the terms on the right side of Eqs.~(\ref{eq:2.8}) that $y^{(2)}(x)$
has the form of Eq.~(\ref{eq:2.6}) and the coefficients $c_k^{(r)}$
can be identified.  Because the equations~(\ref{eq:2.8}) are invariant
under the change of $h$ to $-h$, the coefficients satisfy the relation
$c_{-k}^{(r)}=c_{k}^{(r)}$ for $k=1,2,\dots,r$.  For example, the
first seven sets of coefficients (up to the fifteen-point formula) are
given in Table~\ref{table1}.

\renewcommand{\baselinestretch}{1.25}
\begin{table}[h]
\begin{center}
\begin{tabular}{ccccccccc}
\hline\hline
$r$ & \multicolumn{1}{c}{$k=0$~~~} & \multicolumn{1}{c}{$1$} & 
\multicolumn{1}{c}{$2$} & \multicolumn{1}{c}{$3$} & 
\multicolumn{1}{c}{$4$} & \multicolumn{1}{c}{$5$} & 
\multicolumn{1}{c}{$6$} & \multicolumn{1}{c}{$7$}  \\
\hline
1 & $-2$ & 1 \\
2 & $-\frac{5}{2}$ & $\frac{4}{3}$ & $-\frac{1}{12}$ \\
3 & $-\frac{49}{18}$ & $\frac{3}{2}$ & $-\frac{3}{20}$ & $\frac{1}{90}$ \\
4 & $-\frac{205}{72}$ & $\frac{8}{5}$ & $-\frac{1}{5}$ & $\frac{8}{315}$ 
  & $-\frac{1}{560}$ \\
5 & $-\frac{5269}{1800}$ & $\frac{5}{3}$ & $-\frac{5}{21}$ & $\frac{5}{126}$ 
  & $-\frac{5}{1008}$ & $\frac{1}{3150}$ \\
6 & $-\frac{5369}{1800}$ & $\frac{12}{7}$ & $-\frac{15}{56}$ 
  & $\frac{10}{189}$ & $-\frac{1}{112}$ & $\frac{2}{1925}$ 
  & $-\frac{1}{16632}$ \\
~7~ & $-\frac{266681}{88200}$ & $\frac{7}{4}$ & $-\frac{7}{24}$ 
  & $\frac{7}{108}$ & $-\frac{7}{528}$ & $\frac{7}{3300}$ 
  & $-\frac{7}{30888}$ & $\frac{1}{84084}$ \\
\hline
\end{tabular}
\end{center}
\caption{The coefficients $c_k^{(r)}$ up to $r=7$.}
\label{table1}
\end{table} 
\renewcommand{\baselinestretch}{0.8}

Let us partition the range of $x$ and $t$ values so that
$x_j=x_0+j\Delta x$, $j=0,1,\dots,J$ and $t_n = t_0+n\Delta t$,
$n=0,1,\dots,N$.  The numerical approximation of the wave function at
a mesh point in space and time is denoted as $\psi_{j,n}\approx
\psi(x_j,t_n)$ and we set $V_j = V(x_j)$.  Using
expression~(\ref{eq:2.6}) in Eq.~(\ref{eq:2.5}), we obtain
\begin{eqnarray}
&{\displaystyle
 \psi_{j,n+1}-\frac{i\hbar\Delta t}{4m(\Delta x)^2}\left[
\sum_{k=-r}^{k=r} c_k^{(r)} \psi_{j+k,n+1}\right] + \frac{i\Delta t}{2\hbar} 
V_j\psi_{j,n+1}} & \nonumber  \\
& \mbox{~}\hspace{-0.3in}  {\displaystyle = \psi_{j,n} + \frac{i\hbar\Delta
t}{4m(\Delta x)^2}\left[ \sum_{k=-r}^{k=r} c_k^{(r)}\psi_{j+k,n}\right]
 - \frac{i\Delta t}{2\hbar} V_j\psi_{j,n}}, & 
\label{eq:2.9}
\end{eqnarray}
for $j=0$ to $J$.  The indices in the sums may go out of range, so we
set $\psi_{j,n}=0$ when $j<0$ and $j>J$.  Define
\begin{equation}\label{eq:2.10}
 b \equiv \frac{i\hbar\Delta t}{2m(\Delta x)^2}, \ \ z_1^{(1)} \equiv -2 \ 
\mbox{\rm ~and~} \ a_k^{(r)} \equiv \frac{b}{z_1^{(1)}}c_k^{(r)}, 
\end{equation}
and subsequently
\begin{equation}\label{eq:2.11}
d_j \equiv 1+a_0^{(r)}-\frac{i\Delta t/\hbar}{z_1^{(1)}}V_j, \ \ j=0,1,\dots,J.
\end{equation}
The notation includes $z_1^{(1)}$ which is consistent with that used
in the generalization of the time dependence of the wave function
discussed in the next section.

The solution $\psi_{j,n+1}$ is obtained by solving the system of linear
equations
\begin{equation}\label{eq:2.12}
  A\Psi_{n+1} = A^*\Psi_n,
\end{equation}
where the matrix $A$ is the $(2r+1)$-diagonal matrix
\begin{equation}
A = \left(\begin{array}{cccccccccc}
        d_0 & a_1 & a_2 & \cdots & a_r   & 0 \\
        a_1 & d_1 & a_1 &  \cdots & a_{r-1} & a_r \\
        a_2 & a_1 & d_2 & \cdots & a_{r-2} & a_{r-1} \\
        \vdots & \vdots & \vdots && \vdots & \vdots \\
        a_r & a_{r-1} & a_{r-2} & \cdots & d_r & a_1 \\
        0   & a_r & a_{r-1} & \cdots & a_1 & d_{r+1} \\
            &     &     &     &  && \ddots \\
            &     &     &     &  &&&       & d_{J-1} & a_1 \\
            &     &     &     &  &&&       & a_1     & d_J
        \end{array} \right),
\label{eq:2.14}
\end{equation}
where the superscript $\mbox{}^{(r)}$ of the $a_k$ is assumed.  The
matrix $A^*$ is the complex conjugate of matrix $A$.  The wave
function at $t_{n+1}$, i.e., $\Psi_{n+1}$, is a column vector
consisting of the $\psi_{j,n+1}$ as components, and can be determined
if $\Psi_n$ is known.  The matrix equation (\ref{eq:2.12}) can be
solved using standard techniques.

\section{Time advance}
\label{time}
\setcounter{equation}{0}
In this section we extend the work of Puzynin {\em et
  al.}~\cite{puzynin99,puzynin00}.  The basic idea is to replace the
exponential operator $\exp(-iH\Delta t)$ by the diagonal Pad\'e
approximant.  The $[M/M]$ Pad\'e approximant of the exponential
function may be written as
\begin{equation}\label{eq:3.1}
  f(z) = e^{\textstyle z} = \frac{a_0+a_1z+\cdots + a_Mz^M}{b_0+b_1z
+\cdots + b_Mz^M} 
= \frac{\displaystyle \sum_{m=0}^M a_mz^m}{\displaystyle\sum_{m'=0}^M b_{m'}z^{m'}}, 
\end{equation}
where the $a_m$ and the $b_{m'}$ are complex constants.  It is evident
that when $z=0$, $a_0/b_0 = 1$, which makes one of the coefficients
arbitrary.  By convention we take $b_0=1$ which immediately fixes
$a_0=1$.  There are $2M$ constants remaining, which can be found from
the known coefficients of the series expansion of the exponential
function, giving an error term in Eq.~(\ref{eq:3.1}) ${\cal
  O}(z^{2M+1})$.  The property of Pad\'e approximants that can be used
to advantage is that, if $f(z)$ is unitary, so is its diagonal
Pad\'e approximant~\cite{baker81a}.

In general we solve for the coefficients $a_m$ and $b_{m'}$ by
multiplying Eq.~(\ref{eq:3.1}) by the denominator so that
\begin{equation}\label{eq:3.2}
 \left(\sum_{m'=0}^{M}b_{m'}z^{m'}\right)
 \left(\sum_{i=0}^{\infty}c_iz^i\right)
 = \left(\sum_{m=0}^{M}a_mz^m\right), 
\end{equation} 
where the $c_i$ are known since $e^{\displaystyle z} =
\sum_{i=0}^\infty z^i/i!$.  Multiplying out the sums on the left side
of Eq.~(\ref{eq:3.2}), and equating the coefficients of $z$ through
$z^{2M}$ on both sides, we obtain $2M$ equations in $2M$ unknowns.
The last $M$ of these equations contain no $a_m$ and hence can be
solved for the $b_{m'}$, which in turn can be inserted in the first
$M$ equations to obtain the $a_m$.  The numerator and the denominator
of the diagonal Pad\'e approximant of the exponential function have
been studied extensively~\cite{baker81a}.  When each is factored it is
found that the roots of the denominator are the negative complex
conjugates of the roots of the numerator.  Thus the $[M/M]$ Pad\'e
approximant of the exponential function leads to
\begin{equation}\label{eq:3.3}
    e^{\textstyle z} = \prod_{s=1}^M \left(\frac{1-z/z_s^{(M)}}
      {1+z/\bar{z}_s^{(M)}}
    \right) + {\cal O}(z^{2M+1}),
\end{equation}
where $z_s^{(M)}, \ s =1,\dots,M$, are the roots of the numerator, and
$\bar{z}^{(M)}_s$ is the complex conjugate of $z^{(M)}_s$.  These
roots can be found to a desired precision for virtually any value of
$M$.  We have found them to 17 digit precision for $M$ up to 20, a
sample of which for $M=1$ to 5, each rounded to five decimal places,
is given in Table~\ref{table2}.
\begin{widetext}
\begin{table}[h]
\begin{tabular}{crrrrr}
\hline\hline
$M$ & \multicolumn{1}{c}{$s=1$} & \multicolumn{1}{c}{$2$} 
& \multicolumn{1}{c}{$3$} & \multicolumn{1}{c}{$4$} 
& \multicolumn{1}{c}{$5$}   \\
\hline
1 & $-2.00000+i0.00000$ \\
2 & $-3.00000+i1.73205$ & ~$-3.00000-i1.73205$ \\
3 & $-4.64437+i0.00000$ & ~$-3.67781-i3.50876$ & ~$-3.67781+i3.50876$ \\
4 & $-4.20758+i5.31484$ & ~$-5.79242+i1.73447$ & ~$-5.79242-i1.73446$
  & ~$-4.20758-i5.31483$ \\
5 & $-4.64935+i7.14205$ & ~$-6.70391+i3.48532$ & ~$-7.29348+i0.00000$
  & ~$-6.70391-i3.48532$ & ~$-4.64935-i7.14205$ \\
\hline
\end{tabular}
\protect\caption{The roots $z_s^{(M)}$ of the numerator of the 
    Pad\'e approximant of the exponential function for $M$ from 1 to
    5.}
\label{table2}
\end{table}
\end{widetext}

We use the Pad\'e approximant to express the time evolution operator.
Define the operator
\begin{equation}\label{eq:3.4}
  K_s^{(M)} \equiv \frac{1 - \frac{\textstyle iH\Delta t/\hbar}{\textstyle z_s^{(M)}}}
{1 + \frac{\textstyle iH\Delta t/\hbar}{\textstyle \bar{z}_s^{(M)}}}, 
\end{equation}
so that
\begin{equation}\label{eq:3.5}
  e^{\textstyle -iH\Delta t/\hbar} = \prod_{s =1}^M K_s^{(M)} 
  + {\cal O}((\Delta t)^{2M+1}).
\end{equation}
Since $\Psi_{n+1} = e^{\textstyle -iH\Delta t/\hbar} \Psi_n$, we write the
 relation
\begin{equation}\label{eq:3.6}
  \Psi_{n+1} = \prod_{s=1}^M K_{s}^{(M)}\Psi_n.
\end{equation}
Defining $\Psi_{n+s/M} \equiv K_s^{(M)}\Psi_{n+(s-1)/M}$, we can solve for
$\Psi_{n+1}$ recursively, starting with 
\begin{equation}\label{eq:3.7}
 \Psi_{n+1/M} = K_1^{(M)}\Psi_n. 
\end{equation}  
Assuming that $\Psi_n$ is known, we determine $\Psi_{n+1/M}$
from Eq.~(\ref{eq:3.7}) which has a form similar to that of
Eq.~(\ref{eq:2.5}).  We use therefore the same method of
Sec.~\ref{spatial} to obtain $\Psi_{n+1/M}$.  This is repeated to
obtain in succession
$\Psi_{n+2/M},\Psi_{n+3/M},\dots,\Psi_{n+(M-1)/M},\Psi_{n+1}$.  Since
the operators $K_s^{(M)}$ commute, they can be applied in any order.

\section{Discussion of errors and boundary conditions}
\label{errors}
\subsection{Errors}
\setcounter{equation}{0} In this section we discuss the errors as a
function of the orders of the method, i.e., $r$ and $M$.  Let us
separate the truncation errors due to the integration over space and
those due to integration over time.  At a given time $t$ the spatial
integration with the $r$th-order expansion yields a truncation error
\begin{equation}\label{01}
  e^{(r)} = C^{(r)}(\Delta x)^{2r},
\end{equation}
where $C^{(r)}$ is assumed to be slowly varying with $r$.  Actually
$C^{(r)} = |\psi^{(2r)}(x^*,t)|/(2r!)$ for some $x^*$ in the range of
spatial integration, and thus is model dependent.  If we specify an
acceptable error, the step size $\Delta x$ can be adjusted to obtain
that error.  Since $\Delta x = (x_0 - x_J)/J$, an adjustment of
$\Delta x$ is equivalent to a change in $J$.  Recalling that $x_0-x_J$
is fixed, we obtain
\begin{equation}\label{2}
  \Delta x = \frac{x_0-x_J}{J} = \left(\frac{e^{(r)}}{C^{(r)}}\right)^{1/2r},
\end{equation}
and
\begin{equation}\label{3}
  e^{(r)} \approx \frac{\rm constant}{J^{2r}}.
\end{equation}
We have assumed that $C^{(r)}$ is approximately constant.  The CPU
time for the calculation is proportional to the number of basic
computer operations in solving the matrix equation~(\ref{eq:2.12}).
This involves elementary row operations on $r-1$ rows in $J-1$ columns
to bring the matrix to upper triangular form, plus $J$ back
substitutions to obtain the solution.  Hence
\begin{equation}\label{4}
  \mathrm{CPU~time} \propto \mathrm{\#~operations} \propto Jr \propto
\frac{r}{(e^{(r)})^{1/2r}}. 
\end{equation}
This form gives a minimum (optimum) CPU time which occurs when
\begin{equation}\label{7}
  r \approx -\frac{\ln e^{(r)}}{2}.
\end{equation}

\begin{figure}[htpb]
  \centering
   \resizebox{3.5in}{!}{\includegraphics{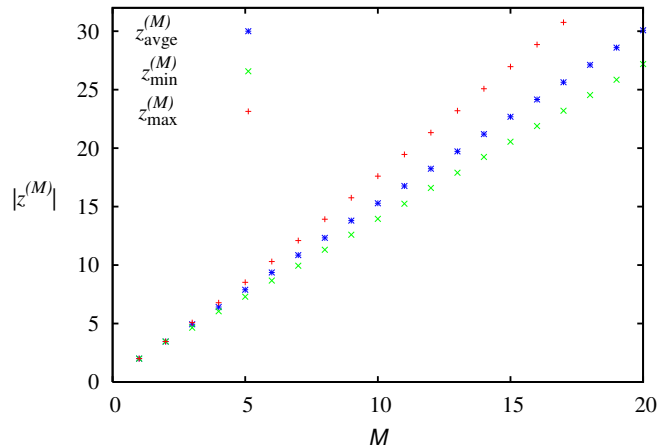}}
   \caption{The average, the minimum and the maximum values of
     $\{|z_s^{(M)}|,s=1\dots M\}$, as a function of $M$.}
\label{figure_2a}
\end{figure}

\begin{figure}[htpb]
  \centering
   \resizebox{3.5in}{!}{\includegraphics{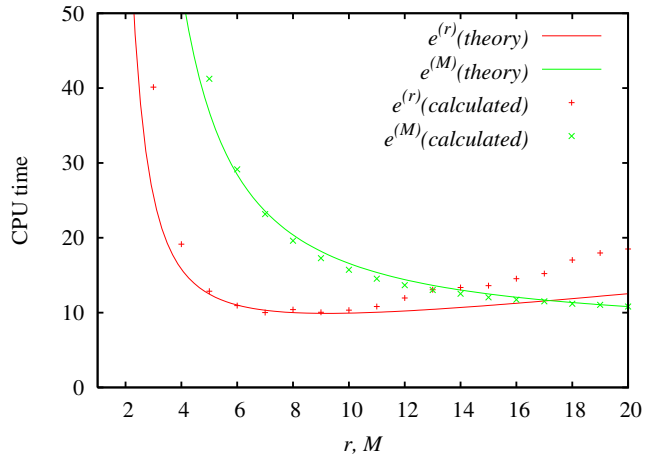}}
   \caption{The normalized theoretical variation of the CPU time for 
     a given error of $1.0\times 10^{-8}$.  The calculated CPU times 
     for the example of Sec.~\ref{oscillation} are shown as dots.}
\label{figure_3a}
\end{figure}

For the time integration we assume a truncation error independent of
$r$.  For a given $r$ the error due to finite $\Delta t$ has a first
term in the expansion
\begin{equation}\label{8}
 e^{(M)} = C^{(M)} (\Delta t)^{2M+1},
\end{equation}
where again $C^{(M)}$ is assumed to be a slowly varying function of
$M$.  We note that the factor $1\over 2$ in the numerator and
denominator of Eq.~(\ref{eq:2.4}) is replaced by $1/z_s^{(M)}$ in each
of the $M$ factors~(\ref{eq:3.4}) of Eq.~(\ref{eq:3.5}).  As $M$
increases the average over different values of $s$ of $|z_s^{(M)}|$,
which we denote as $z^{(M)}_\mathrm{avge}$, also increases.  In fact
$z^{(M)}_\mathrm{avge}$ is a linear function of $M$ as is seen in
Fig.~\ref{figure_2a}.
The effective expansion parameter can be approximated by $2\Delta
t/z^{(M)}_\mathrm{avge}$ rather than $\Delta t$ and hence is
proportional to $\Delta t/M$.  Thus we can replace the relation of
Eq.~(\ref{8}) by
\begin{equation}\label{9}
 e^{(M)} \approx C^{(M)} (\Delta t/M)^{2M+1},
\end{equation}
where the constant $C^{(M)}$ is appropriately adjusted.  If we take
the total time $t_\mathrm{max} = N\Delta t$ to be fixed, then
\begin{equation}\label{10}
  \mathrm{CPU~time} \propto \left(e^{(M)}\right)^
  {-\frac{\textstyle 1}{\textstyle 2M+1}}.
\end{equation}
In Fig.~\ref{figure_3a} the curves of the (scaled) CPU times are
plotted.  Both curves clearly show the sharp decline when $r,M$
increases from 1 through 5 or larger.  For increasing $M$ the CPU time
continues to decline although the decrements become smaller at larger
$M$.  For increasing $r$ there is a minimum depending on the specified
error and beyond the minimum the curve shows a slow increase with
increasing $r$.  Superimposed on the curves are the CPU times (as
dots) of a model calculation (see Sec.~\ref{oscillation}), in which
the numerical and exact solutions can be compared.  Clearly the
theoretical trends, including the minimum as a function of $r$, occur
in the computed example.
It should be noted that it ``pays'' to increase $M$ indefinitely,
whereas there is an optimum value of $r$ which depends on the
magnitude of $e^{(r)}$.

\subsection{Boundary conditions}
\label{bc}
Below Eq.~(\ref{eq:2.9}) we indicate that we set $\psi_{j,n} =0$ when
$j<0$ or $j>J$, or when $j$ goes out of range.  These are appropriate
boundary conditions when the wave function and its first $r+1$
derivatives are zero at the boundaries, since it was assumed in the
derivation of the method that all these derivatives exist.  If however
that is not the case, for instance, at the boundary of an (in)finite
square well or barrier where the second-order derivative does not
exist, one must devise ways of incorporating the proper boundary
conditions.

One case of importance, which we discuss in the third example (see
Sec.~\ref{decay}) of the paper, is the case of radial behavior of a
partial wave when angular momentum decomposition has been done.  In
the $S$-wave case the wave function, defined only for nonnegative
values of the radial coordinate, is zero at the origin but the first
derivative is finite.  Muller~\cite{muller99} discusses a related, but
not identical, situation.  He considers three-point formulas for the
Coulomb potential which lead to a radial wave function which is zero
when the radial variable $\rho=0$, but has first and second
derivatives which are nonzero at $\rho=0$.  His approach can be
adapted to the $(2r+1)$-point formula of this work.

We treat this case by making the ansatz that the wave function behaves
like an odd function about the origin and continues in the unphysical
region of the negative radial variable.  With this assumption we do
not affect the behavior of the system at positive values of the
radial variable, but all the required derivatives exist.  Furthermore,
the wave function at negative $j$ values can be combined with the ones
with corresponding positive $j$ values, so that the space need not be
enlarged but instead the first few matrix elements of the matrix $A$
can be changed to account for the boundary condition.  This is
achieved by replacing $A$ by $A' = A-B$ in Eq.~(2.13) where
\begin{equation}\label{eq:12}
B=\left(
\begin{array}{cccccccccc}
0 & a_1 & a_2 & a_3 & \cdots & a_{r-2} & a_{r-1} & a_r & 0 & \cdots \\
0 & a_2 & a_3 & a_4 & \cdots & a_{r-1} & a_r &  0 & 0 & \cdots \\
0 & a_3 & a_4 & a_5 & \cdots & a_r & 0 & 0 & 0 & \cdots \\
\vdots & \vdots & \vdots & \vdots && \vdots & \vdots & \vdots & \vdots &  \\
0 & a_{r-1} & a_r & 0 & \cdots & 0 & 0 & 0 & 0 & \cdots \\
0 & a_r & 0 & 0 & \cdots & 0 & 0 & 0 & 0 & \cdots \\
0 & 0 & 0 & 0 & \cdots & 0 & 0 & 0 & 0 & \cdots \\
\vdots & \vdots & \vdots & \vdots &&   \vdots & \vdots & \vdots & \vdots &  \\
\end{array} \right).
\end{equation}

A hard-core type potential could be dealt with in the same way.
Different forms of boundary conditions are more complicated to
implement, but Ref.~\cite{muller99} suggests an approach to including
such boundary conditions.  For the purpose of the radial wave function
of a nonsingular potential, the above approach is sufficient.  

\section{Examples}
\label{examples}
\setcounter{equation}{0} We consider three systems to which this
numerical method may be applied.  The first two allow us to make a
comparison with the exact solution and to test the precision of the
numerical procedure.  The third involves the time evolution of a
quasi-stable quantum process.

\subsection{Oscillation of a coherent wavepacket}
\label{oscillation}
The oscillation of a coherent state in the harmonic oscillator well is
described in Ref.~\cite{schiff68}.  The time evolution of
such states has been discussed recently in connection with the quantum
abacus.  (See Ref.~\cite{cheon04}.)  In that case there is a point
interaction at the center of the oscillator well.  It is of interest
to consider narrow but finite-range interactions to simulate more
realistic physical systems.  In order to test the robustness of the
quantum gates one needs a very stable numerical procedure.  We test
the precision of the numerical procedure by investigating the case
without the central interaction, so that the numerical results can be
compared with the exact ones.

The time-dependent Schr\"odinger equation is
\begin{equation}\label{eq:4.1}
  i\hbar\frac{\partial~}{\partial t} \psi(x,t) = \left(-\frac{\hbar^2}{2m}
\frac{\partial^2~}{\partial x^2} + \frac{1}{2}Kx^2\right) \psi(x,t).
\end{equation}
We consider the time evolution of the initially displaced ground-state
wave function
\begin{equation}\label{eq:4.2}
  \psi(x,0) = \frac{\alpha^{1/2}}{\pi^{1/4}}e^{-{1\over 2}\alpha^2(x-a)^2},
\end{equation}
where $\alpha^4 = mK/\hbar^2$, $\omega = \sqrt{K/m}$, and $a$ is the
initial displacement.
The closed expression for the time evolved wave function is
\begin{eqnarray}
\psi_\mathrm{exact} & = & \frac{\alpha^{1/2}}{\pi^{1/4}}\exp\left[-\frac{1}{2}
(\xi - \xi_0\cos{\omega t})^2 \right. \nonumber \\
&& \left. -i(\frac{1}{2} + \xi\xi_0\sin{\omega t} 
-\frac{1}{4}\xi_0^2\sin{2\omega t})\right],
\label{eq:4.3}
\end{eqnarray}
where $\xi=\alpha x$ and $\xi_0 = \alpha a$.  We set $\hbar = m = 1$,
$\omega = 0.2$, and $a = 10$.  We choose our space such that
$x\in[x_0,x_J] = [-40,40]$.  The period of oscillation is then $T =
10\pi$.  We allow the coherent state to oscillate for eleven periods
before comparing the numerical solution to the exact one.  The error
is calculated as $e_2$ using the formula~\cite{puzynin99}
\begin{equation}\label{eq:4.4}
  (e_2)^2 = \int_{x_0}^{x_J} \ dx \ |\psi(x,t_1)-\psi_\mathrm{exact}(x,t_1)|^2,
\end{equation}
where $t_1 = 11T$ for our example.  The results including the relative
CPU time~\protect\footnote{The CPU time is a relative measure.  The
  calculations of Table~\ref{table3} were done on a computer with an
  Intel Pentium 4 Mobile CPU 1.90 Ghz processor.  Those of
  Table~\ref{table4} were done with a AMD Opteron processor 250
  running at 2.4 GHz.  The CPU times within a table give a rough
  comparison of the efficiency of the method.  CPU times in different
  tables should not be compared.}  are displayed in Table~\ref{table3}.
\begin{table}[htb]
\begin{tabular}{ccccccc}
\hline
~$M$~ & ~$r$~ & ~$\Delta t$~ & ~$\Delta x$~ & ~~$J$~~ & $e_2$ & CPU time \\
\hline\hline
20 & 20 & $\pi$ & 0.44444 & 180 & $6.717\times 10^{-9}$ & 18.38 \\ 
20 & 15 & $\pi$ & 0.38095 & 210 & $7.044\times 10^{-9}$ & 14.23 \\ 
20 & 10 & $\pi$ & 0.27586 & 290 & $7.506\times 10^{-9}$ & 10.84 \\ 
20 & 7  & $\pi$ & 0.18182 & 440 & $9.353\times 10^{-9}$ & 10.12 \\
20 & 5 & $\pi$ & 0.09877 & 810 & $9.330\times 10^{-9}$ & 12.83 \\ 
20 & 4 & $\pi$ & 0.05755 & 1390 & $9.871\times 10^{-9}$ & 18.61 \\ 
20 & 3 & $\pi$ & 0.03810 & 2100 & $2.102\times 10^{-7}$ & 23.67 \\ 
20 & 2 & $\pi$ & 0.03810 & 2100 & $1.624\times 10^{-4}$ & 18.42 \\ 
20 & 1 & $\pi$ & 0.03810 & 2100 & $1.749\times 10^{-1}$ & 13.75 \\ 
\hline 20 & 10 & $\pi$ & 0.26667 & 300 & $5.106\times 10^{-9}$ & 10.77 \\ 
15 & 10 & $\pi/1.5$ & 0.26667 & 300 & $5.153\times 10^{-9}$ & 12.13 \\ 
10 & 10 & $\pi/3$ & 0.26667 & 300 & $4.995\times 10^{-9}$ & 16.16 \\ 
5 & 10 & $\pi/15$ & 0.26667 & 300 & $8.787\times 10^{-9}$ & 40.42 \\ 
3 & 10 & $\pi/150$ & 0.26667 & 300 & $1.840\times 10^{-9}$ & 242.9 \\ 
1 & 10 & $\pi/3000$ & 0.26667 & 300 & $5.046\times 10^{-4}$ & 1627 \\ 
\hline
\end{tabular}
\caption{Summary of computational time and errors incurred by using 
  the numerical integration procedure when the initial wave function 
  is the displaced ground state.  The last column indicates a relative 
  CPU run time.  The upper half of the table gives the effects of 
  changing the number of spatial steps; the lower half the effects of 
  changing the number of time steps.} 
\label{table3}
\end{table}

In the above tests we have tried to obtain a precision better than
$10^{-8}$.  While varying the number of steps for the spatial
integration, we kept the number of time factors per time step constant
at 20.  Given that the total space is fixed and spans 80 units, we
adjusted the number of spatial steps $J$ to give the required
precision.  We limited (arbitrarily) the maximum number of spatial
steps to 2100.  With $M=20$ the 15-point formula ($r=7$) is most
efficient.  When $r<4$ (less than 9-point formula), we were unable to
reach the precision criterion because of the imposed limit on $J$. It
is clear from the trend however that the efficiency is significantly
less for the lower $r$ values.  The 9-point formula is roughly half as
efficient as the 15-point formula.

The effect of different order time formulas as seen in the lower part
of Table~\ref{table3} is even more dramatic.  For the spatial
integration we used the 21-point formula ($r=10$), and varied the
time-order formula, i.e., $M$, from 20 to 1.  We see at least two
orders of magnitude improvement in computational speed as $M$ is
increased over this range.

A comparison with the standard CN approach ($r=M=1$) is instructive.
We considered the same system with $x_0=-25$ and $x_J=25$, $\Delta x =
0.005$ and $\Delta t = 0.5(\Delta x)^2$.  The standard CN method
yielded an error of $e_2=7.1\times10^{-5}$ when $t=T/4$ which
increased exponentially to $e_2=2.7\times 10^{-3}$ at $t=10T$.  Whereas
the CPU time in Table~\ref{table3} is given in seconds, the CPU time
required to complete this last calculation exceeded 24 hours.

The computed CPU times shown in Fig.~\ref{figure_3a} exceeded the
``theoretic'' values by increasing amounts as $r$ increased beyond 10.
This can be attributed to the approximate nature of the error analysis
in which the model dependence of $C^{(r)}$ (and $C^{(M)}$) was
neglected.  In this example a more elaborate analysis could be done
since the wave function is known analytically.  In practical
situations where a numerical method is used the analytic wave function
is usually not known and an estimate such as we have given here would
be all that is available.  The main point is that dramatic
improvements result both theoretically and computationally when larger
values of $r$ and $M$ are employed.

\subsection{Propagation of a wavepacket}
\label{wavepacket}
For this example we return to the work of Ref.~\cite{goldberg67} and
consider the main features of that analysis with a view of determining
the improvement brought about by the generalizations of this paper.
This problem was revisited by Moyer~\cite{moyer04} to illustrate the
efficacy of the Numerov method and the use of transparent boundary
conditions for the propagation of free-particle wavepackets.  The
authors of Ref.~\cite{goldberg67} consider wavepackets impinging on a
square barrier and study their behavior in time.  We consider first
free wavepacket propagation (without potential), and second the
reflection and transmission of a wavepacket by a smooth potential.

Thus we first assume $V(x)=0$ and take as initial wave function
\begin{equation}\label{eq:4.5}
  \psi(x,0) = (2\pi\sigma_0^2)^{-1/4}
    e^{\textstyle ik_0(x-x_0)}e^{\textstyle -(x-x_0)^2/(2\sigma_0)^2}.
\end{equation}
(Note that our $\sigma_0$ is that of Ref.~\cite{goldberg67} divided by
$\sqrt{2}$.)  The wave function at later time is given by
\begin{displaymath}
\mbox{~\hspace{-0.75in}}\psi(x,t)  =  (2\pi\sigma_0^2)^{-1/4}
  [1+i\hbar t/(2m\sigma_0^2)]^{-1/2} \times  \nonumber
\end{displaymath}
\begin{equation}\label{eq:4.6}
 \exp{\left\{\frac{\textstyle -(x-x_0)^2/(2\sigma_0)^2 + 
      ik_0(x-x_0)-i\hbar k_0^2t/(2m)}{\textstyle 1 + i\hbar t/(2m\sigma_0^2)}
  \right\}}.
\end{equation}

We use parameters comparable to those of Ref.~\cite{goldberg67}.  We
set $\hbar = 1$ and $m={1\over 2}$.  The coordinate range we take is
from $-0.5$ to 1.5 rather than from 0 to 1 since over the smaller space
the normalization of the packet is not as precise as we require
because the tails of the Gaussian are nonzero outside the (0,1)
interval.  We choose $\sigma_0 = 1/20$, $k_0=50\pi$, $\Delta t =
2(\Delta x)^2$, and allow as much time as it takes the packet
to travel from $x_0 = 0.25$ to around 0.75.  For the final position
the numerically calculated wave function is compared to the analytic
one and $e_2$ is determined.  In Table~\ref{table4}
\begin{table}[hb]
\begin{tabular}{ccccccc}
\hline
~$M$~ & ~$r$~ & ~~$J$~~ & $e_2$ & CPU time \\
\hline\hline
1 & 1 &  2000 & $9.418\times 10^{-2}$ & 2.20 \\ 
  &   &  4000 & $2.189\times 10^{-2}$ & 18.04 \\ 
  &   &  8000 & $5.368\times 10^{-3}$ & 151.92 \\ 
  &   &  16000 & $1.336\times 10^{-3}$ & 1287.8 \\
2 &	2 &	2000 &		$3.018\times 10^{-4}$ &	5.99 \\
3 &	3 &	2000 &		$1.321\times 10^{-6}$ &	12.57 \\
4 &	4 &	2000 &		$6.577\times 10^{-9}$ &	22.64 \\
5 &	5 &	2000 &		$3.648\times 10^{-11}$ & 37.13 \\
6 &	6 &	2000 &		$8.437\times 10^{-13}$ & 56.05 \\
10 & 10  &  440 & $3.606\times 10^{-9}$ & 2.12 \\ 
20 & 20  &  260 & $4.542\times 10^{-9}$ & 2.51 \\
\hline
\end{tabular}
\caption{Summary of computational parameters used to calculate the 
propagating free packet and compare it to the analytic wavepacket.}
\label{table4}
\end{table}
we list some of the computed results.  We observe that the traditional
CN method ($M=1$ and $r=1$) has a low precision and that using greater
$J$ (smaller $\Delta x$) results in modest gain in precision.  By
using higher-order time formula one can make significant gain in
precision (seven orders of magnitude) with no increase in
computational time compared with that of Ref.~\cite{goldberg67}.
(Compare the first row to the last two rows of Table~\ref{table4}.)
Rows 5 through 9 of Table~\ref{table4} illustrate the transition from
less precise to more precise solutions.  It is consistent with the
finding of the authors of Ref.~\cite{misicu01} who use an $M=2$, $r=3$
method and obtain two orders of magnitude improvement of the results
of Refs.~\cite{bertulani99,vandijk99a}.  The results are sensitive to
surprisingly high orders of $\Delta x$ and $\Delta t$.

\begin{figure}[htpb]
  \centering
   \resizebox{3.5in}{!}{\includegraphics{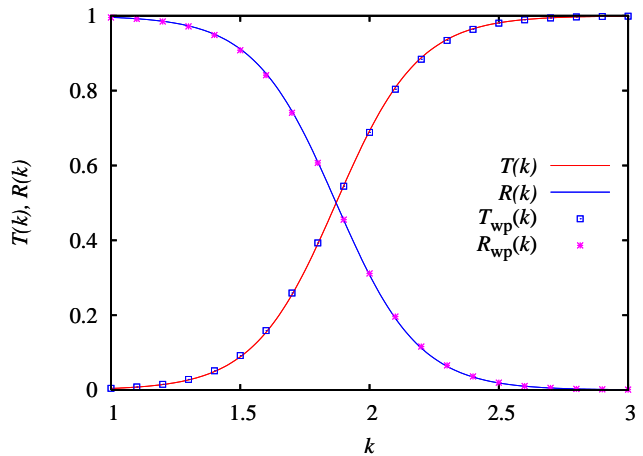}}
\caption{The transmission and reflection coefficients as a function of
  $k$ when $\beta=1$, $\lambda=2.5$, $m=1$, and $\sigma_0=10$.  The
  subscript ``wp'' indicates that the coefficients are obtained from
  the emerging wavepackets.  The quantities without subscripts are
  calculated using the time-independent method.}
\label{figure_1}
\end{figure}

Another test using wavepacket scattering to show the efficacy of the
higher-order approach is scattering from a potential.  Rather than
using the square barrier of Ref.~\cite{goldberg67}, we consider the
repulsive P\"oschl-Teller type potential~\cite{flugge74,landau77} of
the form
\begin{equation}\label{eq:4.7}
  V(x) = \frac{\hbar^2}{2m}\frac{\beta^2\lambda(\lambda-1)}
  {\cosh^2 \beta x}.
\end{equation}   
Since this potential does not have discontinuities the improved CN
method works well with it.  The transmission and reflection
coefficients are known analytically.  We can also compute them by
considering the wavepacket Eq.~(\ref{eq:4.5}) incident on the
potential.  Over a sufficiently long time the wavepacket will have
interacted with the potential and transmitted and reflected packets
emerge and travel away from the potential region.  At that point we
can calculate the probabilities of the particle represented by the
packet on the left and on the right of the potential; these
probabilities correspond to the transmission and reflection
coefficients provided the packet is sufficiently narrow in momentum
space.  This means that one needs an initial packet which is wide in
coordinate space.  In our calculation we choose $\beta = 1$, $\lambda
= 2.5$, $m = 1$, and $\sigma_0 = 10$.  This gives a spread in the
incident momentum-space wave packet of $\sigma_k = 0.05$.  The width
in momentum space of the reflected and transmitted wavepackets also
has this value.  The domain of the $x$ coordinates is from $-300$ to
$+300$ and the initial position of the packet is at $x_0=-150$ to
ensure that there is no overlap of the initial packet and the
potential.  We find good agreement between the transmission and
reflections probabilities determined by plane-wave scattering approach
and the time-dependent calculation as shown in Fig.~\ref{figure_1}.

\subsection{Long-time behavior of decay of quasi-stable system}
\label{decay}
There are few analytically solvable models of the time evolution of
unstable quantum systems~\cite{vandijk02,vandijk03}.  Realistic
systems need to be solved numerically.  Long-time calculations are
required for systems which require both nuclear and atomic time-scales
such as ionization and bremsstrahlung due to radioactive decay of the
nucleus of an atom~\cite{kataoka00,bertulani99,misicu01}.  To study the
short-time anomalous power law behavior of the decay and the
long-time inverse-power law behavior the method of this paper is
appropriate.  This is especially relevant because of the recently
observed violation of the exponential-decay law at long
times~\cite{rothe06}.

\begin{figure}[htpb]
  \centering
   \resizebox{3.5in}{!}{\includegraphics{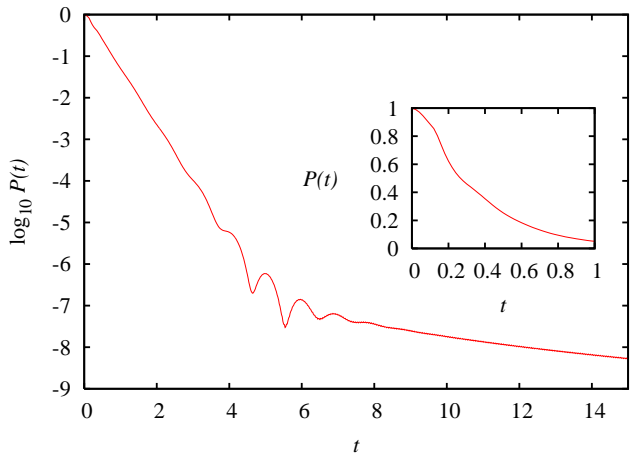}}
\caption{The nonescape probability as a function of time for the 
  interaction with $\lambda = 3$, $a = 1$, and $w=0.10$.  We also take
  $\hbar = 1$ and $m = {1\over 2}$.}
\label{figure_2}
\end{figure}

To illustrate the numerical method discussed in this paper as applied
to decaying systems let us consider a variant of the model with a
$\delta$-shell potential~\cite{winter61}, but with the $\delta$
function replaced by a gaussian.  Thus in the $S$ partial wave of a
spherical system the potential is
\begin{equation}\label{eq:4.8}
  V(\rho) = \frac{\lambda}{w\sqrt{\pi}}\exp{[-(\rho-a)^2/w^2]}, 
\end{equation}
where $\rho$ is the radial coordinate.  This potential reduces to the
$\delta$-shell potential, $V_\delta(\rho)=\lambda\delta(\rho-a)$ when
$w\rightarrow 0$.  For small but finite values of $w$ this potential
leads to scattering results which are good approximations of those of
the $\delta$-shell interaction~\protect\footnote{To be published.}.
Initially the quantum system is in the state
\begin{equation}\label{eq:4.9}
  \psi(\rho,0) = \sqrt{2/a}\sin{(\pi \rho/a)}.
\end{equation}
In our example we take $\hbar = 1$, $\lambda = 3$, $m = {1\over 2}$,
$a=1$ and $w = 0.10$.  Using the numerical method of this paper,
including the modification of matrix $A$ as described in Sec.~\ref{bc}
to take care of the boundary conditions at $\rho=0$, we determine the
wave function at later times, i.e., $\psi(\rho,t)$.  From that we
obtain the nonescape probability, as a function of $t$, $P(t) =
\int_0^a |\psi(\rho,t)|^2 \ d\rho$, which is shown in
Fig.~\ref{figure_2}.
It clearly shows the exponential decay-region in time, the inverse
power-law behavior for long times, the deviation from exponential
decay at short times, and the transition regions~\cite{winter61}.

\begin{figure}[htpb]
  \centering
   \resizebox{3.5in}{!}{\includegraphics{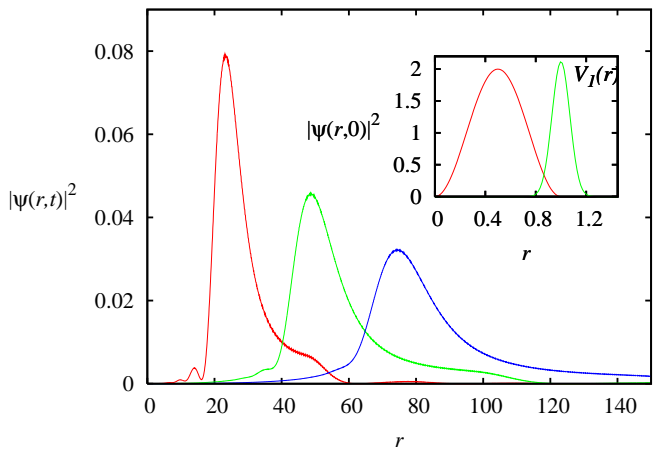}}
\caption{The square of the absolute value of the wave function as a 
  function of $\rho$ at times $t=5, 10, 15$ for the same parameters apply
  as in Fig.~\ref{figure_2}.  The insert gives the $t=0$ graph as well
  as the scaled potential, $V_1(\rho)=V(\rho)/8$.}
\label{figure_3}
\end{figure}

The quadratic short-time behavior is seen in Fig.~\ref{figure_2} as is
the inverse power law behavior at long times~\cite{khaflin58}.
Remarkably the decaying system can be studied in this manner for a
time exceeding thirty half-lives.  In Fig.~\ref{figure_3} we plot the
square of the absolute value of the wave function at times $t = 5, 10,
15$.
Notice that the wave function (packet) has three distinct regions: a
precursor due to energy components of the initial wave function larger
than that associated with the exponential decay, the main packet which
corresponds to the exponential decay at the resonance energy, and the
follower, which is a small blip that stretches in time (travels
more slowly) and is due to energy components in the initial wave function
which have lower energy that the resonance energy.  If the maximum
spatial coordinate, which is set at 800 for this calculation, were set
at a smaller value, say 400, then one observes a fuzziness in the
precursor of the right-most wave function.  This can be attributed to
the finite space in which the wavepacket travels so that the fast
precursor has been partially reflected from the right boundary and
interferes with the wave front of the main wavepacket.  The numerical
parameters for this calculation are $\Delta\rho = 0.1$, $\Delta t =
0.02$, $r = 20$ and $M = 20$. 


\section{Remarks}
\label{conclusions}
The generalized CN method that we have presented in this paper gives
many orders of magnitude improvement in the precision of the results
and several orders of magnitude in the computational time required to
obtain the results.  Clearly, since this method enhances the
efficiency of the numerical calculations, it can be a significant tool
for studying time-dependent processes.  It goes beyond the improvement
of Ref.~\cite{misicu01} in a systematic way.  It also is an advance
over the method of Ref.~\cite{moyer04}, since the Numerov method has
an error ${\cal O}((\Delta x)^6)$, and it is difficult to see how it
can be generalized systematically to higher order spatial errors.  The
generalized time evolution algorithm can be applied to
Moyer's~\cite{moyer04} method.

We have applied the method for $r$ and $M$ up to and including 20 for
both.  Having achieved significant improvements with these values of
$r$ and $M$ we did not consider larger values although there does not
seem to be a practical reason that this cannot be done.  Although the
approach seems to saturate at $r$ around 10 (see Table~\ref{table3}),
there is no visible saturation in the time-evolution part of the
problem.  One would expect higher orders of spatial errors to be
significant when the wave function and/or the potential has large
spatial fluctuations.  Even so, at $r=7$ we are using a 15-point
formula for the second-order spatial derivative, and it is surprising
that a smaller number of points in the formula is not sufficient for
optimal results in this case.  It should be noted that the higher
order method as discussed in this paper are suitable only for
well-behaved, sufficiently differentiable solutions; these occur when
the potential function is well behaved.  As the authors of
Ref.~\cite{press_C92} point out for singular functions higher-order
methods do not necessarily lead to greater accuracy.
 
It should be noted that in this paper we consider primarily
one-dimensional systems, but the method applies equally well to
partial-wave equations of two- or three-dimensional systems.  The
study of the decaying quasi-stable state is an example of the latter.

An interesting avenue to investigate further is the impact that this
approach may have on two- or three-dimensional systems, where the
number of variables involved is equal to the dimension.  The
Peaceman-Rachford-type approach~\cite{peaceman55}, also known as the
alternating-direction implicit method, of factoring the approximation
of the time evolution operator may apply as it did in
Refs.~\cite{galbraith84,kulander82} or more recently in, for example,
Refs.~\cite{shon00,ishikawa04}.  Using the Crank-Nicolson method the
authors of Refs.~\cite{galbraith84,kulander82} show that in two
dimension the kinetic energy parts of the evolution operator
factorizes.  Whether such factorization can be generalized in the
spirit of the method of this paper is under investigation.
  
Calculations on one-dimensional multichannel systems indicate that
this approach also leads to substantially greater efficiencies.

Preliminary studies with the Numerov spatial integration
scheme~\cite{moyer04} and the generalized time evolution as described
in this work indicate that significant improvements occur if one
incorporates appropriate changes in the spatial step size for
different regions of space.  This is important in the case of
discontinuous potentials and potentials that have great variation in
some region and little or no variation in other regions.  Furthermore
it is well known that the wavepacket has large fluctuation in a
(short-range) potential region and little variation in the asymptotic
regions.  A great savings in computational time can be achieved by
using different space-step sizes in the different regions.  One needs
to investigate whether such variable step size can be incorporated in
the generalized spatial integration scheme of this paper.  The
approach that dealt with the discontinuous first- or second-order
derivative in this paper and Ref.~\cite{muller99} is worth exploring.
We intend to study this in the future.

\acknowledgments

We are grateful to Professor Y. Nogami for carefully reading the
manuscript and the constructive comments, as well as useful and
helpful discussions.  One of the authors (WvD) expresses gratitude for
the hospitality of the Department of Information and Communication
Sciences of Kyoto Sangyo University where most of this work was
completed.  He also acknowledges the financial support for this
research from the Natural Sciences and Engineering Council of Canada
and the Japan Society for the Promotion of Science.


\end{document}